# AssemblyNet: A large ensemble of CNNs for 3D Whole Brain MRI Segmentation

Pierrick Coupé[1], Boris Mansencal[1], Michaël Clément[1], Rémi Giraud[2], Baudouin Denis de Senneville[3], Vinh-Thong Ta[1], Vincent Lepetit[1] and José V. Manjon[4]

[1] CNRS, Univ. Bordeaux, Bordeaux INP, LABRI, UMR5800, F-33400 Talence, France
[2] Bordeaux INP, Univ. Bordeaux, CNRS, IMS, UMR 5218, F-33400 Talence, France
[3] CNRS, Univ. Bordeaux, IMB, UMR 5251, F-33400 Talence, France
[4] ITACA, Universitat Politècnica de València, 46022 Valencia, Spain

## Abstract

*Whole brain segmentation using deep learning (DL) is a very challenging task since the number of anatomical labels is very high compared to the number of available training images. To address this problem, previous DL methods proposed to use a single convolution neural network (CNN) or few independent CNNs. In this paper, we present a novel ensemble method based on a large number of CNNs processing different overlapping brain areas. Inspired by parliamentary decision-making systems, we propose a framework called AssemblyNet, made of two "assemblies" of U-Nets. Such a parliamentary system is capable of dealing with complex decisions, unseen problem and reaching a consensus quickly. AssemblyNet introduces sharing of knowledge among neighboring U-Nets, an "amendment" procedure made by the second assembly at higher-resolution to refine the decision taken by the first one, and a final decision obtained by majority voting. During our validation, AssemblyNet showed competitive performance compared to state-of-the-art methods such as U-Net, Joint label fusion and SLANT. Moreover, we investigated the scan-rescan consistency and the robustness to disease effects of our method. These experiences demonstrated the reliability of AssemblyNet. Finally, we showed the interest of using semi-supervised learning to improve the performance of our method.*

## 1 Introduction

Quantitative brain analysis is crucial to better understand the human brain and to analyze different brain pathologies. However, whole brain segmentation is still a very challenging problem, mostly due to the high number of anatomical labels compared to the limited number of available training data. Manual segmentation of the whole brain is indeed a very tedious and difficult task, preventing the production of large annotated datasets.

To address this question, several methods have been proposed in the past years. By extending the single-atlas method paradigm, the multi-atlas framework has been successfully applied to whole brain segmentation [1], [2]. In such approaches, labeled templates are first nonlinearly registered to the target image. Afterwards, the estimated deformations are applied to the manual segmentations before fusing them. This type of methods efficiently deals with limited training data, however, the required multiple nonlinear registrations can result in a huge computational time. Moreover, regularization involved in registration may prevent to accurately capture complex local anatomical patterns.

To reduce the computational time of multi-atlas methods and to better capture local anatomy, patch-based methods have been introduced [3]. In such approaches, the label fusion step is based on the nonlocal patch-based estimator. These methods demonstrated state-of-the-art performance for whole brain segmentation [4]–[6][7]. One of the main references in the domain is the patch-based joint label fusion (JLF) which won the MICCAI challenge in 2012 [4] and which is still considered as the state of the art for whole brain segmentation. In patch-based methods, usual machine learning such as sparse coding [8] or neural networks [9] has been used in place of the nonlocal estimator. Recently, a fast framework has been proposed [10] to further reduce the computational time required by patch-based methods.

More recently, deep leaning (DL) methods have also been proposed for whole brain segmentation. Due to limited GPU memory, first attempts were based on patchwise strategies [11]–[13] or 2D segmentation (slice by slice) [14][15][16]. Only recently, first 3D fully convolutional network methods were proposed using reduced input data size (*i.e.*, 128×128×128 voxels) [17] or using Spatially Localized Atlas Network Tiles (SLANT) strategy [18]. This latter framework divides the whole volume into overlapping sub-volumes, each one being processed by a different U-Net [19] (*e.g.*, 8 or 27). The SLANT strategy addresses the problem of limited GPU memory and simplifies the complex problem of whole brain segmentation into simpler problems, better suited to limited training data.

In this paper, we propose to extend this framework by using a much larger number of simpler 3D U-Nets (*i.e.*, 250) while keeping processing time similar. The main question to address is the optimal organization of this large ensemble of CNNs. To this end, we propose a new framework we call AssemblyNet. Inspired by the decision-making process developed by human societies to deal with complex problems, we decided to model a parliamentary system based on two separate assemblies. Such bicameral – meaning two chambers – parliament has been adopted by many countries around the world. A bicameral system is usually composed of an upper and a lower chamber, both having their own independency to ensure the balance of power. However, an assembly may communicate its vote to the other for amendment. Such parliamentary system is capable of dealing with complex decisions, unseen problem and reaching a consensus quickly. This study extends our conference paper [20] with more complete experiments investigating *i)* the impact of semi-supervised learning *ii)* the scan-rescan reliability of our method and *iii)* the robustness to disease effects of the proposed AssemblyNet.

## 2  Materials and Methods

### 2.1  Method overview

In AssemblyNet, both assemblies are composed of 3D U-Nets considered as "assembly members" (see Figure 1). Each member represents one territory (*i.e.*, brain area) in the final vote. To this end, we used spatially localized networks where each U-Net only processes a sub-volume of the global volume, as done in [18]. Sub-volumes overlap each other, so the final segmentation results from an overcomplete aggregation of local votes. A majority vote is used to obtain the global segmentation. Moreover, each member can share knowledge with their nearest neighbor in the assembly. In particular, we propose a novel nearest neighbor transfer learning strategy, where weights of the spatially nearest U-Net are used to initialize the next U-Net.

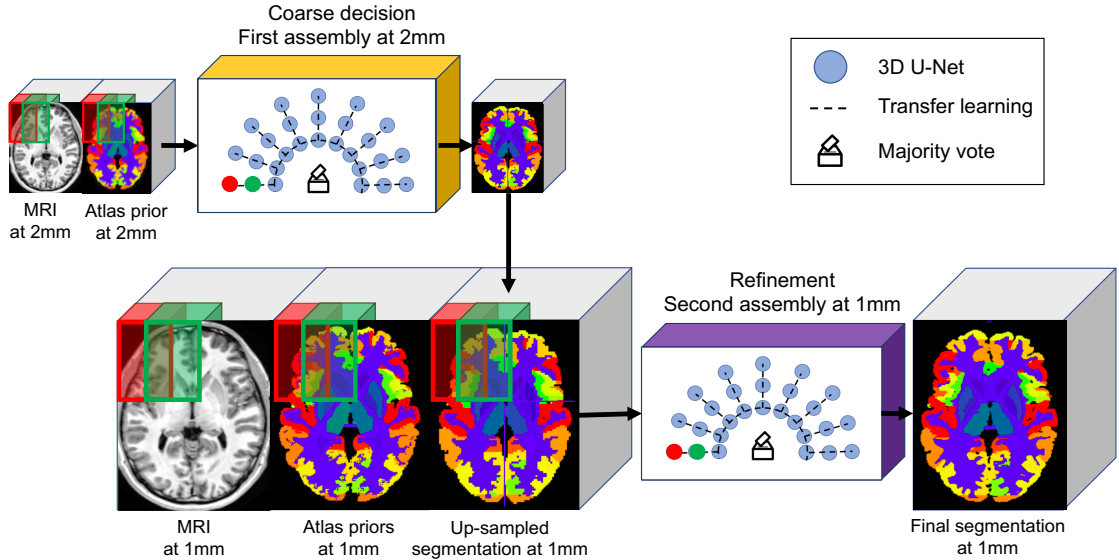

*Figure 1: Illustration of the proposed AssemblyNet framework. Our method is based on two assemblies of 125 3D U-Nets integrated into a multiscale framework. The first assembly (in yellow) provides a coarse segmentation at 2×2×2 mm³. The second assembly (in purple) refines this coarse segmentation to produce the final segmentation at 1×1×1 mm³. Each 3D U-Net processes a different but overlapping area of the brain. The U-Nets in red in both Assemblies process the area indicated by red rectangles in the input images. The U-Nets in green process the area indicated by green rectangles. During training, the U-Nets in green are initialized using weights of the U-Nets in red by transfer learning. The output segmentations for each assembly are obtained by majority voting of the 125 3D U-Nets.*

In addition, we also propose to use prior knowledge on the expected final decision which can be viewed as the bill (*i.e.*, draft law) submitted to an assembly for consideration. As prior knowledge, we decided to use nonlinearly registered Atlas prior.

Finally, we also propose modeling communication between both assemblies using an innovative multiscale strategy. In AssemblyNet, we use a multiscale cascade of assemblies where the first assembly produces a coarse decision at 2×2×2 mm³. This coarse decision is transmitted to the second assembly for analysis at 1×1×1 mm³. This amendment procedure is similar to an error correction or a refinement step. After consideration by both assemblies, the bill under consideration becomes a law which represents the final segmentation in our system.

Our contributions are: *i)* the use of prior knowledge based on fast atlas registration, *ii)* a knowledge sharing between CNNs using nearest neighbor transfer learning, *iii)* iterative refinement process based on a multiscale cascade of assemblies and *iv)* the use of student-teacher semi-supervised learning based on a well-designed auxiliary dataset.

## 2.2 Datasets

**Training dataset**: 45 T1w MRI from the OASIS dataset [21] manually labeled according to the BrainCOLOR protocol were used for training. The selected images were the same than the ones used in [18]. The age range was 18-96y for this dataset. All the used images and manual segmentations were obtained from Neuromorphometrics Inc. During our experiments, we used the 132 anatomical labels consistent across subjects (see [18]).

**Testing dataset**: 19 T1w MRI manually labeled according to the BrainCOLOR protocol were used for testing. These MRI came from three different datasets: 5 from the OASIS dataset, one from the colin27 cohort [22] and 13 from the CANDI database [23]. This testing dataset is the same one used in [18]. The age range was 20-89y for OASIS, 5-15y for CANDI and the age was 27y for Colin27.

**Scan-Rescan dataset:** 8 T1w MRI (from 4 subjects) manually labeled according to the BrainCOLOR protocol were used for scan-rescan experiment. The same expert segmented both the scan and the rescan image. These MRI came from 2 different datasets: 3 from the OASIS dataset (not included in the training) and one from a patient with Alzheimer's disease from the ADNI dataset [24].

**Pathological dataset:** 29 T1w MRI manually labeled according to the BrainCOLOR protocol were included in the pathological dataset. We did not use the rescan image of the AD patient described in the scan-rescan dataset. These MRI came from the ADNI dataset. There were 15 cognitively control subjects and 14 patients with Alzheimer's disease in this dataset. The age of the subjects varied from 62y to 88y.

**Lifespan dataset:** 360 unlabeled T1w MRI were randomly selected under constraints from the dataset used in our previous BigData studies [25], [26] to build the lifespan dataset. This dataset was based on 9 datasets publicly available (C-MIND[1], NDAR[2], ABIDE[3], ICBM[4], IXI[5], OASIS[6], AIBL[7], ADNI1 and ADNI2[8]). From 1y to 90y, we selected 2 females and 2 males for each age (*i.e.*, 2F and 2M of 1 year old, 2F and 2M of 2 years old and so on). Therefore, we obtained a balanced group with 50% of each gender uniformly distributed from 1y to 90y. We made sure that none of the training or testing subjects were selected in this auxiliary dataset.

## 2.3 AssemblyNet framework

**Preprocessing**: To homogenize input orientations and intensities, all the images were first preprocessed using the volBrain pipeline[9] [27] with the following steps: *i)* denoising [28], *ii)* inhomogeneity correction [29], *iii)* affine registration into the MNI space (181×217×181 voxels at 1×1×1 mm$^3$) [2], *iv)* tissue-based intensity normalization [30] and *v)* brain extraction [31]. Finally, image intensities were centralized and normalized within the brain mask.

**Atlas priors:** To obtain priors knowledge on the expected results, we performed a nonlinear registration of a MICCAI 2012 Multi-Atlas Labeling Challenge atlas (based on the 45 training images) to the subject under consideration. The Atlas prior was built using Ants [2] and majority voting. The non-inear registration of the Atlas prior to the subject under study was done with an unsupervised deep learning framework similar to VoxelMorph trained on the lifespan dataset [32].

**Assembly description:** Each assembly was composed of 125 3D U-Nets equally distributed in the MNI space along each axis (*i.e.,* 5 along x, y and z). We experimentally found that 5×5×5 to produced better results than 4×4×4 and similar to 6×6×6. Each 3D U-Net processed a sub-volume large enough to ensure at least 50% of overlap between sub-spaces. At the end, a majority vote was used to aggregate the local votes.

**Nearest neighbor transfer learning:** To enable knowledge sharing between U-Nets of an assembly, we propose a new transfer learning where the weights of the nearest U-Net were used to initialize the next U-Net. In practice, we only copied the weights of the descending path of the U-Net architecture. At the beginning, we trained the first U-

---

[1] https://research.cchmc.org/c-mind/
[2] https://ndar.nih.gov
[3] http://fcon_1000.projects.nitrc.org/indi/abide/
[4] http://www.loni.usc.edu/ICBM/
[5] http://brain-development.org/ixi-dataset/
[6] http://www.oasis-brains.org
[7] http://adni.loni.ucla.edu/research/protocols/mri-protocols
[8] www.loni.usc.edu
[9] https://www.volbrain.upv.es

Net from scratch. Then, each U-Net on the first column was initialized with weights of the previous U-Net. Once the first column was trained, each U-Net of the next column was initialized with the U-Net at the same position on the previous column and so on. Finally, once the first 2D plane of U-Nets was trained, each U-Net of the next 2D planes was initialized with the U-Net at the same position on the previous plane and so on.

**Multiscale cascade of assemblies:** To make our decision-making system faster and more robust, we decided to use a multiscale framework. Consequently, the first assembly at 2×2×2mm$^3$ produced a coarse segmentation. Afterwards, an up-sampling of this segmentation to 1×1×1mm$^3$ was performed using nearest neighbor interpolation. The second assembly estimated the final result at 1×1×1mm$^3$.

## 2.4 Semi-supervised learning framework

In this study, we also investigated the use of semi-supervised learning (SSL) to further improve segmentation accuracy. SSL aims at using a small amount of labeled data in combination with a large number of unlabeled images to achieve higher performance. Therefore, SSL is particularly interesting in medical images analysis where the manual segmentations of experts are limited. In [14], [18], authors proposed to use auxiliary libraries segmented with traditional tools such as Freesurfer [33] or non-local spatial staple label fusion (NLSS) [7] to improve their whole brain segmentation framework based on deep learning. However, such methods can require impractical computational burden (*e.g.*, 21 CPU years in [18]) and classical tools may provide suboptimal auxiliary segmentations.

Here, we propose to follow the teacher-student paradigm [34] to take advantage of the fast processing capabilities and high segmentation accuracy of the proposed AssemblyNet. The teacher-student paradigm has been efficiently used for image classification task [35]. In our SSL framework, an AssemblyNet teacher – trained on the 45 training images – was first used to segment unlabeled images (*i.e.*, the 360 images of the lifespan dataset). Then, these 360 pseudo-labeled images were used to train from scratch an AssemblyNet student. At the end, the AssemblyNet student was fine-tuned on the 45 manual segmentations of the training dataset. During our experiments, we investigated the iteration of this procedure considering that the obtained AssemblyNet student could be a good teacher for a second student generation. In our SSL framework, we took care to build the unlabeled images dataset balanced in age and gender in order to limit bias introduction in the pseudo-labeled population.

## 2.5 Implementation details

**Data augmentation**: First, the images of the training and lifespan datasets were flipped along mid sagittal plane in the MNI space. Then, we used MixUp data augmentation during training to minimize overfitting problems [36]. This method performs a linear interpolation of a random pair of training examples and their corresponding labels.

**Training framework:** For all the networks, we used the 3D U-Net architecture proposed in [18], but with a lower number of filters. Instead of using a basis of 32 filters of 3×3×3 – 32 for the first layer, 64 for the second and so on – we selected a basis of 24 filters of 3×3×3 to reduce by 25% the network size. We experimentally found that this setting reduced memory consumption without impacting performance. Moreover, we used the same parameters for all the U-Nets with: batch size = 1, optimizer = Adam, epoch = 100, loss = Dice and dropout = 0.5 after each block of the descending path. For the U-Nets of the first assembly at 2×2×2 mm$^3$, we used input resolution = 32×48×32 voxels and input channel = 2 (*i.e.*, T1w and Atlas priors). For the U-Nets of the second assembly at 1×1×1 mm$^3$, we used input resolution = 64×72×64 voxels and input channel = 3 (*i.e.*, T1w, Atlas priors and up-sampled coarse segmentation). In addition, to compensate for the small batch size, we performed temporal averaging of model weights [37]. At the end of the 100 epochs, we performed additional 20 epochs where the model estimated at each epoch is averaged with previous ones using a moving average. Such average of model weights along the optimization trajectory leads to better generalization than usual training [38]. For the SSL step, we used only 20 epochs for normal optimization and 10 epochs for moving average. Finally, we also performed dropout at test time [39]. For each U-Net, we generated 3 different outputs before averaging them (with dropout layer active). Such method helps reducing variance of the networks. As in [18], the experiments were done with an NVIDIA Titan Xp with 12 GB memory and thus processing times are comparable.

**Computational time:** The preprocessing steps take around 90s. The nonlinear registration of the atlas takes less than 5s thanks to a deep leaning framework similar to [32]. The first assembly at 2×2×2 mm$^3$ requires 3min to segment an image while the second assembly at 1×1×1 mm$^3$ requires 5min. At the end, the final segmentation is registered back to the native space using the inverse affine transform estimated during preprocessing. This interpolation takes around 30s. Therefore, the full AssemblyNet process takes around 10min including preprocessing, segmentation, and inverse registration back to the native space.

## 2.6 Validation framework

First, for each testing subject, we estimated the average Dice coefficient on the 132 considered anatomical labels (without background) in the native space. Afterwards, we estimated the global mean Dice in % over the 19 images of the testing dataset. In this experiment, we compared AssemblyNet with several state-of-the-art methods. First, the patch-based joint label fusion (JLF) [4] was used as reference. In addition, we included U-Net [19], SLANT-8 and SLANT-27 methods as proposed in [18]. SLANT-8 is based on 8 U-Nets processing non-overlapping sub-volumes of 86×110×78 voxels while SLANT-27 is based on 27 U-Nets processing overlapping sub-volumes 96×128×88 voxels. All these methods were trained on the same 45 training images described in the section datasets. Finally, we included SLANT-27 trained on 5111 auxiliary images segmented using NLSS [7] and fined tuned on the 45 training images. These are the best published results for whole brain segmentation to our knowledge. For all these methods we report the results published in [18]. We also used the docker implementation of SLANT-27[10] to produce segmentations of all the considered datasets.

For the scan-rescan reliability experiment, we rigidly registered the re-scan image into the native space of the scan images [2]. We then interpolated the re-scan segmentations into the native space of the scan images using the estimated transformation matrix. By this way, we estimated the Dice coefficients between both manual segmentations (*i.e.*, intra-rater consistency) and both automatic segmentations (*i.e.*, intra-method consistency). Moreover, we estimated the method-expert consistency as the Dice coefficients between the automatic segmentation of the rescan images and the manual segmentation of the scan image.

For the experiment on the robustness to disease effects, we first computed the Dice on the 29 ADNI subjects. We then compared the Dice coefficients obtained for cognitively normal (CN) subjects and patients with Alzheimer's Disease (AD) to study the impact of pathology on segmentation accuracy.

For all these experiments, we used one-sided non-parametric Wilcoxon signed-rank test at 95% of confidence to assess the significance of Dice improvement as in [18]. Moreover, we used one-sided Mann-Whitney rank test at 95% of confidence to assess the significance of Dice decrease between the AD group compared to the CN group.

---

[10] https://github.com/MASILab/SLANTbrainSeg

# 3 Results

## 3.1 AssemblyNet performance

First, we evaluated the proposed contributions (see Table 1). Compared to baseline results at 2×2×2 mm$^3$ (Dice=67.4%), the use of Atlas prior provided a gain of 0.3 percentage point (pp) in terms of mean Dice. Moreover, the combination of Atlas prior and transfer learning improved by 0.5 pp the baseline mean Dice. In addition, multiscale cascade of assemblies increased by 1.1 pp the mean Dice obtained with Assembly at 1×1×1 mm$^3$ without multiscale cascade (Dice=72.2%). Finally, AssemblyNet outperformed by 5.9 pp the mean Dice obtained with baseline Assembly at 2×2×2 mm$^3$. The Dice coefficients produced by AssemblyNet were significantly better than the Dice coefficients produced by all the considered alternatives. Note these results were obtained using only 45 training cases.

*Table 1: Evaluation of the proposed contributions. The mean Dice is evaluated on the 19 images of the test dataset in the native space for the 132 considered labels (without background). Testing time includes image preprocessing and registration back to the native space. \* indicates a significant lower Dice compared to AssemblyNet using a Wilcoxon test.*

| Methods | Atlas prior | Transfer learning | Multi-scale | Dice in % | Training time | Testing time |
|---|---|---|---|---|---|---|
| **Assembly at 2×2×2 mm$^3$** | No | No | - | 67.4 (3.4)* | 29h | 5min |
| **Assembly at 2×2×2 mm$^3$** | Yes | No | - | 67.7 (3.3)* | 29h | 5min |
| **Assembly at 2×2×2 mm$^3$** | Yes | Yes | - | 67.9 (3.3)* | 29h | 5min |
| **Assembly at 1×1×1 mm$^3$** | Yes | Yes | No | 72.2 (3.8)* | 6 days | 7min |
| **AssemblyNet** | Yes | Yes | Yes | **73.3 (4.2)** | 7 days | 10min |

## 3.2 Impact of semi-supervised learning

Second, we evaluated the impact of the proposed teacher-student semi-supervised learning (SSL) framework (see Table 2). The used of the lifespan dataset – labeled with teacher – in the training of the student lead to an improvement of 0.6pp of mean Dice compared to the mean Dice of the teacher (Dice=73.3%). The fine-tuning (FT) step further increased the mean Dice by 0.3 pp. These results are in line with previous literature on the role of the FT step [18], [35]. Finally, the second iteration with FT produced marginal improvement and lead to a mean Dice of 74.0%. This improvement was not significant. Therefore, in the following, we used the first student generation since the time required for the second iteration is not justified by the performance improvement.

*Table 2: Impact of the proposed semi-supervised learning framework on the 19 images of the testing dataset. The mean Dice is evaluated on the 132 considered labels (without background) in the native space. \* indicates a significant lower Dice compared to second generation of AssemblyNet with semi-supervised learning and fine tuning (i.e., 360 + FT 45) using a Wilcoxon test.*

| Methods | Training images | Dice in % | Training time | Library extension time |
|---|---|---|---|---|
| **AssemblyNet** Teacher | 45 | 73.3 (4.2)* | 7 days | 0s |
| **AssemblyNet** First student generation | 360 | 73.6 (4.1)* | 12 days | 2.5 days |
| **AssemblyNet** First student generation | 360 + FT 45 | 73.9 (4.0) | 14 days | 2.5 days |
| **AssemblyNet** Second student generation | 360 | 73.9 (4.0) | 26 days | 5 days |
| **AssemblyNet** Second student generation | 360 + FT 45 | 74.0 (3.9) | 28 days | 5 days |

## 3.3 Comparison with state-of-the-art methods

We compared AssemblyNet with state-of-the-art methods (see Table 3). When considering only methods trained with 45 images, AssemblyNet improved mean Dice obtained with U-Net and SLANT-8 by 16.3 pp, JLF by 9.9 pp and SLANT-27 by 7.2 pp. AssemblyNet was also efficient in term of training and testing times compared to SLANT-based methods. It has to be noted that Assembly at 2×2×2 mm outperformed all the methods using 45 training images (except AssemblyNet) while working at lower resolution.

Table 3: Comparison with state-of-the-art methods on the 19 images of the testing dataset. The mean Dice is evaluated on the 132 considered labels (without background) in the native space. * indicates a significant lower Dice compared to AssemblyNet with semi-supervised learning (i.e., 360 + FT 45) using a Wilcoxon test.

| Methods | Training images | Dice in % | Training time | Testing time | Library extension time |
|---|---|---|---|---|---|
| **U-Net [19]** | 45 | 57.0 | 33h | 8min | 0s |
| **SLANT-8 [18]** | 45 | 57.0 | 11 days | 10min | 0s |
| **JLF [4]** | 45 | 63.4 | 0s | 34h | 0s |
| **SLANT-27 [18]** | 45 | 66.1 | 42 days | 15min | 0s |
| **SLANT-27 [18]** | 5111 + FT 45 | 72.9 | 27 days | 15min | 21 years [a] |
| **Assembly at 2×2×2 mm** | 45 | 67.9 (3.3)* | 29h | 5min | 0s |
| **SLANT-27 (docker)** | 5111 + FT 45 | 72.6 (2.8)* | 27 days | 15min | 21 years [a] |
| **AssemblyNet** | 45 | 73.3 (4.2)* | 7 days | 10min | 0s |
| **AssemblyNet** | 360 + FT 45 | **73.9 (4.0)** | 14 days | 10min | 2.5 days |

[a] Library extension time represents the CPU time required to segment 5111 MRI using NLSS (*i.e.*, 34h×5111). This number of 21 CPU years is reported in [18].

Compared to SLANT-27 trained over 5111+45 images, our method provided better results without library extension while being faster to train and to execute. Using our SSL framework based on 360+45 images, AssemblyNet obtained a gain of 1pp compared to SLANT-27 trained over 5111+45 images. According to [18], their library extension required 21 CPU years to be completed. Consequently, such an approach is impractical or very costly using a cloud-based solution. The proposed SSL framework is more practical in term of time and resources.

Finally, our AssemblyNet using SSL was significantly better than AssemblyNet without SSL and SLANT-27 (docker). In our framework, SLANT-27 (docker) obtained slightly lower results than the ones published in the original paper [18]. This may come from hardware and environment differences.

In addition, we analyzed the performance of the methods according to the dataset. Mean Dice coefficients obtained on each testing dataset (*i.e.*, OASIS, CANDI and Colin27) are provided in Table 4. As expected, all the methods performed better on adult scans from the OASIS dataset since the training dataset comes from the same cohort. Moreover, all the images were acquired with the same protocol on the same scanner for this dataset. First, we can note the good performance of JLF compared to U-Net and SLANT-8. Moreover, when considering methods trained with 45 training, AssemblyNet outperformed U-Net by 8.2 pp, JLF by 4.4 pp and SLANT-27 by 1.2 pp of mean Dice. When considering all the methods, AssemblyNet with SSL obtained significantly better Dice than SLANT-27 (docker) trained on 5111+45 images. It has to be noted that in [18], the authors have shown that the SLANT-27 method significantly outperformed U-Net, JLF and SLANT-8. Finally, on the OASIS images, using SSL did not significantly improve the AssemblyNet results.

Table 4: Comparison with state-of-the-art methods on the different testing datasets (5 adult scans from OASIS, 13 child scans from CANDI child and the high-resolution Colin27 image based on scans average). Using Wilcoxon tests * indicates a significant lower Dice compared to AssemblyNet with semi-supervised learning (i.e., 360 + FT 45) when compared to SLANT-27 (docker) and AssemblyNet.

| Methods | Training images | OASIS Dice in % | CANDI Dice in % | Colin27 Dice in % |
| --- | --- | --- | --- | --- |
| **U-Net [19]** | 45 | 70.6 (0.9) | 51.4 (8.1) | 62.1 |
| **SLANT-8 [18]** | 45 | 69.9 (1.4) | 51.9 (7.0) | 59.7 |
| **JLF [4]** | 45 | 74.6 (0.9) | 59.0 (3.3) | 64.6 |
| **SLANT-27 [18]** | 45 | 76.6 (0.8) | 62.1 (6.2) | 66.5 |
| **SLANT-27 [18]** | 5111 + FT 45 | 77.6 (1.2) | 71.1 (2.3) | 73.2 |
| **SLANT-27** (docker) | 5111 + FT 45 | 75.9 (1.7)* | 71.3 (2.2)* | 73.5 |
| **AssemblyNet** | 45 | 78.8 (1.7) | 71.1 (2.9)* | 74.2 |
| **AssemblyNet** | 360 + FT 45 | **79.0** (2.0) | **71.9** (2.9) | **75.0** |

On child scans from the CANDI dataset acquired with a different protocol, we can first note a dramatic drop in performance for all the methods except for AssemblyNet with and without SSL and SLANT-27 trained on 5111+45 images. Moreover, when considering methods trained with 45 training, AssemblyNet outperformed U-Net by 19.7 pp, JLF by 19.2 pp and SLANT-27 by 9 pp of mean Dice. This demonstrates the robustness and the generalization capability of the proposed framework with regards to unseen acquisition protocol and ages. When considering all the methods, AssemblyNet with SSL obtained significantly better Dice than AssemblyNet without SSL and SLANT-27 (docker).

On the high-resolution Colin27 image, we also observed an important decrease of performance for all the methods except for AssemblyNet and SLANT-27 trained on 5111+45 images. As for CANDI dataset, AssemblyNet obtained the best segmentation accuracy with or without SSL on this dataset.

As an illustration, Figure 2 shows the segmentations of the central slides in the native space obtained by SLANT-27 (docker) and AssemblyNet with SSL on the first subject of the testing dataset (ID=1120_3). Both methods provided good segmentations although AssemblyNet segmentation was less smooth especially around sulci (*e.g.*, cerebellum – see red ellipses). Moreover, we can observe an over segmentation of cortical gray matter in SLANT-27 segmentation as visible in the error map where structures appeared (see green ellipses). Finally, this figure shows the staircasing artifacts present in the human segmentation (*e.g.*, pallidum – see the pink ellipses) while automatic methods were more regular and consistent.

### 3.4 Scan-rescan consistency

The study of segmentation reproducibility produced by a segmentation method is also highly important especially in medical imaging. Therefore, we carried out a scan-rescan experiment to investigate the consistency and reliability of the proposed method. The results on the 4 images of the scan-rescan images are provided in Table 5. It has to be highlighted that there are variations between both acquisitions due to patient's motion, distortion, inhomogeneity and noise. Therefore, agreement between both acquisitions is not expected to be perfect but higher Dice indicates better method stability, consistency and reliability.

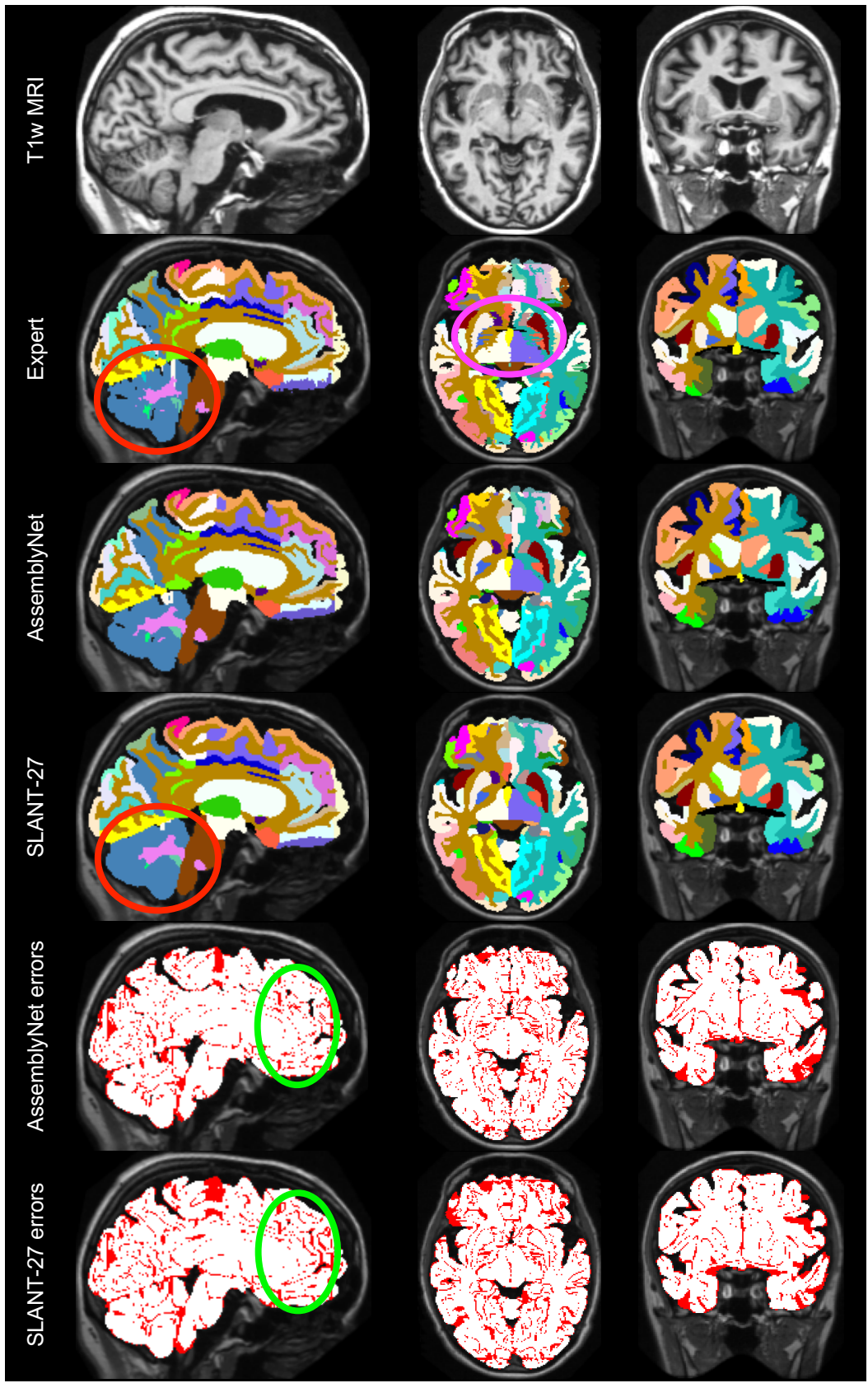

*Figure 2: Example of segmentations in the native space for the first testing subject (ID=1120_3). First rows: sagittal; axial and coronal views for the T1w MRI. Second row: manual segmentation produced by the expert. Third row: segmentation obtained by our AssemblyNet with SSL. Fourth row: segmentation obtained by SLANT-27 (docker). Fifth row: binary difference between manual and AssemblyNet segmentations. Last row: binary difference between manual and SLANT-27 (docker) segmentations. Colored ellipses indicate areas of interest*

*Table 5: Reliability study on the scan-rescan datasets (3 adult scans from OASIS and one scan of a patient with AD from ADNI). The intra-method consistency is the mean Dice between the automatic segmentations obtained on the scan and rescan images. The Expert-Method consistency is the mean Dice between the automatic segmentation on the rescan image and the manual segmentation of the scan image. The Dice coefficients obtained on the scan and the rescan images are averaged. The intra-rater consistency is the mean Dice between the manual segmentations obtained on the scan and rescan images. Using Wilcoxon tests * indicates a significant lower Dice compared to AssemblyNet with semi-supervised learning (i.e., 360 + FT 45).*

| Methods | Training images | Intra-method consistency | Intra-rater consistency | Expert-Method consistency |
|---|---|---|---|---|
| **SLANT-27** (docker) | 5111 + FT 45 | 91.7 (1.5)* | | 72.9 (1.7)* |
| **AssemblyNet** | 45 | 92.0 (1.8)* | 76.8 (4.1) | 75.6 (1.7) |
| **AssemblyNet** | 360 + FT 45 | **92.8 (1.8)** | | **75.8 (1.9)** |

First, we estimated the consistency of the segmentations provided by SLANT-27 (docker), AssemblyNet without SSL, AssemblyNet with SSL and the expert on the scan and the rescan images. As expected, we can note that automatic methods were much more consistent (>90% of Dice) than the expert who obtained a scan-rescan segmentation consistency of 76.8% of Dice. Moreover, we can see that AssemblyNet was more consistent than SLANT-27 especially when using the proposed teacher-student SSL framework. The method consistency of AssemblyNet with SSL was significantly higher than the SLANT-27 (docker) and AssemblyNet without SSL.

In addition, we estimated the Expert-Method consistency as the mean Dice coefficients between the automatic segmentation on the rescan image and the manual segmentation of the scan image. The Expert-Method consistency of AssemblyNet with SSL was significantly higher than the consistency of SLANT-27 (docker) but not than the consistency of AssemblyNet without SSL. Finally, the Expert-Method consistencies obtained by automatic methods were not significantly lower than the intra-expert consistency although this difference was almost significant for SLANT-27 docker (p=0.36 for AssemblyNet with SSL, p=0.23 for AssemblyNet without SSL and p=0.07 for SLANT-27).

### 3.5 Robustness to disease effects

The last part of our validation is dedicated to robustness to disease effects. To this end, we compared SLANT-27 (docker) and AssemblyNet on the pathological dataset composed of CN subjects and AD patients.

First, we estimated the mean Dice for both groups (see Table 6). For the CN group, AssemblyNet with SSL obtained a significantly better Dice than both other methods. For the AD group, we obtained similar results although the improvement obtained by AssemblyNet with SSL was higher for AD group (2.2 pp) than for the CN group (1.7 pp) compared to SLANT-27 (docker).

In addition, we compared the accuracy between CN and AD groups for the three methods. We found no significant differences between groups for all the methods although this difference was almost significant for SLANT-27 docker (p=0.31 for AssemblyNet with SSL, p=0.18 for AssemblyNet with SSL and p=0.06 for SLANT-27).

Finally, AssemblyNet obtained a global Dice 73.1% without SSL and 73.6% with SSL, while SLANT-27 trained on 5111+45 obtained 71.6%. The results of AssemblyNet with SSL were significantly better than the results obtained with both other methods.

Table 6: Methods comparison on the pathological dataset (29 scans from the ADNI dataset including 15 CN and 14 patients with AD). * indicates a significant lower Dice compared to AssemblyNet with semi-supervised learning (i.e., 360 + FT 45) using Wilcoxon tests.

| Methods | Training images | Dice in % on CN | Dice in % on AD | ADNI Dice in % |
|---|---|---|---|---|
| **SLANT-27** (docker) | 5111 + FT 45 | 72.3 (1.6)* | 71.0 (2.6)* | 71.6 (2.2)* |
| **AssemblyNet** | 45 | 73.6 (1.6)* | 72.6 (2.6)* | 73.1 (2.2)* |
| **AssemblyNet** | 360 + FT 45 | **74.0 (1.5)** | **73.2 (2.5)** | **73.6 (2.1)** |

## 4  Discussion

In this work, we presented a novel whole brain segmentation framework based on a large number of 3D CNN (*i.e.*, 250 U-Nets) called AssemblyNet. First, we showed that the use of Atlas prior, nearest neighbor transfer learning and multiscale cascade of Assemblies enable to improve global segmentation accuracy. In further work alternative options could be investigated. First, atlas prior could be replaced by fast multi-atlas prior. Thanks to nonlinear registration methods based on deep learning, such prior is no more too expensive. Moreover, more advanced communication between assembly members should be investigated. Recent advances in multi-agent reinforcement learning seem a promising way [40]. Finally, in this paper, we focused only on the optimal organization of a large group of CNNs without studying the optimal assembly composition. Additional works should investigate this point, for instance by introducing model diversity in the assembly.

Second, we studied the impact of the proposed SSL based on a teacher-student paradigm. We showed that using few hundreds of well-balanced unlabeled data could significantly improve the results of AssemblyNet in all the cases (*i.e.,* unseen acquisition protocol, age period and pathology). Compared to previous methods using larger auxiliary datasets labelled with classical tools [14], [18], the proposed SSL framework is more practical in term of computational time and resources. However, SSL is currently receiving special attention in deep learning community. Consequently new paradigms should be considered [41], [42].

Afterwards, we compared our AssemblyNet with state-of-the-art methods. We demonstrated the high performance of our method in terms of segmentation accuracy and computational time. First, these experiments demonstrated the advantage of using several CNNs to segment the whole brain since SLANT-27 and AssemblyNet clearly outperformed the use of a single U-Net. Moreover, these results showed that using a larger number of simpler CNNs within a multiscale framework is an efficient strategy. While for OASIS we obtained Dice higher than intra-expert consistency, the accuracy on the CANDI dataset is still limited. This can come from several factors such as the lower image quality (*e.g.,* more motion artifacts in child images) or the larger distance between adult training dataset and this child dataset. These points should be deeper investigated in future whole brain segmentation methods. In terms of computational time, our method could be further improved by using several GPUs. At training time, once the first U-Net is trained, several following U-Nets can be trained in parallel despite our transfer learning strategy. At testing time, AssemblyNet can be fully parallelized and thus the processing time could be drastically reduced using multiple GPUs.

In addition, we investigated the scan-rescan consistency of the proposed method. We showed that the intra-method consistency of our method reached 92.8% while intra-rater consistency was limited to 76.8%. This result clearly demonstrates that our automatic method segments the whole brain in a more consistent manner than a human expert. Moreover, we show that the expert-method consistency obtained with our method is not significantly lower than the intra-rater consistency, which is an encouraging result. However, these results were obtained using only 4 scan-rescan subjects. In addition, the Morphometric dataset does not contain material to evaluate method reproducibility (same image segmented twice by the same expert). Such data would have been useful to evaluate if automatic methods had reached human variability. Finally, these results raise the question of using human segmentation as "gold standard" with a consistency lower than 80%. Semi-manual segmentations could be considered in order to reduce intra-rater variability.

Finally, we studied the robustness of AssemblyNet to pathology. To this end, we compared its accuracy on CN and AD groups. We observed a small but non-significant decrease of Dice for the AD group compared to CN group. Moreover, compared to SLANT-27, AssemblyNet was less impacted by the presence of the pathology. This is a first step towards a more extensive validation with other pathologies.

## 5 Conclusion

In this paper, we proposed to use a large number of CNNs to perform whole brain segmentation. We investigated how to organize this large ensemble of CNNs to quickly and accurately segment the brain. To this end, we designed a novel deep decision-making process called AssemblyNet based on two assemblies of 125 3D U-Nets. Our validation showed the very competitive results of AssemblyNet compared to state-of-the-art methods. We also demonstrated that AssemblyNet is very efficient to deal with limited training data and to accurately achieve segmentation in a practical training and testing times. Finally, we demonstrated the interest of semi-supervised learning to improve the performance of our method on unseen acquisition protocol, age period and pathology.

# Acknowledgements

This work benefited from the support of the project DeepvolBrain of the French National Research Agency (ANR-18-CE45-0013). This study was achieved within the context of the Laboratory of Excellence TRAIL ANR-10-LABX-57 for the BigDataBrain project. Moreover, we thank the Investments for the future Program IdEx Bordeaux (ANR-10-IDEX- 03- 02, HL-MRI Project), Cluster of excellence CPU and the CNRS/INSERM for the DeepMultiBrain project. This study has been also supported by the DPI2017-87743-R grant from the Spanish Ministerio de Economia, Industria Competitividad. The authors gratefully acknowledge the support of NVIDIA Corporation with their donation of the TITAN Xp GPU used in this research.

Moreover, this work is based on multiple samples. We wish to thank all investigators of these projects who collected these datasets and made them freely accessible.

The C-MIND data used in the preparation of this article were obtained from the C-MIND Data Repository (accessed in Feb 2015) created by the C-MIND study of Normal Brain Development. This is a multisite, longitudinal study of typically developing children from ages newborn through young adulthood conducted by Cincinnati Children's Hospital Medical Center and UCLA and supported by the National Institute of Child Health and Human Development (Contract #s HHSN275200900018C). A listing of the participating sites and a complete listing of the study investigators can be found at https://research.cchmc.org/c-mind.

The NDAR data used in the preparation of this manuscript were obtained from the NIH-supported National Database for Autism Research (NDAR). NDAR is a collaborative informatics system created by the National Institutes of Health to provide a national resource to support and accelerate research in autism. The NDAR dataset includes data from the NIH Pediatric MRI Data Repository created by the NIH MRI Study of Normal Brain Development. This is a multisite, longitudinal study of typically developing children from ages newborn through young adulthood conducted by the Brain Development Cooperative Group and supported by the National Institute of Child Health and Human Development, the National Institute on Drug Abuse, the National Institute of Mental Health, and the National Institute of Neurological Disorders and Stroke (Contract #s N01- HD02-3343, N01-MH9-0002, and N01-NS-9-2314, -2315, -2316, -2317, -2319 and -2320). A listing of the participating sites and a complete listing of the study investigators can be found at http://pediatricmri.nih.gov/nihpd/info/participating_centers.html.

The ADNI data used in the preparation of this manuscript were obtained from the Alzheimer's Disease Neuroimaging Initiative (ADNI) (National Institutes of Health Grant U01 AG024904). The ADNI is funded by the National Institute on Aging and the National Institute of Biomedical Imaging and Bioengineering and through generous contributions from the following: Abbott, AstraZeneca AB, Bayer Schering Pharma AG, Bristol-Myers Squibb, Eisai Global Clinical Development, Elan Corporation, Genentech, GE Healthcare, GlaxoSmithKline, Innogenetics NV, Johnson & Johnson, Eli Lilly and Co., Medpace, Inc., Merck and Co., Inc., Novartis AG, Pfizer Inc., F. Hoffmann-La Roche, Schering-Plough, Synarc Inc., as well as nonprofit partners, the Alzheimer's Association and Alzheimer's Drug Discovery Foundation, with participation from the U.S. Food and Drug Administration. Private sector contributions to the ADNI are facilitated by the Foundation for the National Institutes of Health (www.fnih.org). The grantee organization is the Northern California Institute for Research and Education, and the study was coordinated by the Alzheimer's Disease Cooperative Study at the University of California, San Diego. ADNI data are disseminated by the Laboratory for NeuroImaging at the University of California, Los Angeles. This research was also supported by NIH grants P30AG010129, K01 AG030514 and the Dana Foundation.

The OASIS data used in the preparation of this manuscript were obtained from the OASIS project funded by grants P50 AG05681, P01 AG03991, R01 AG021910, P50 MH071616, U24 RR021382, R01 MH56584. See http://www.oasis-brains.org/ for more details.

The AIBL data used in the preparation of this manuscript were obtained from the AIBL study of ageing funded by the Common-wealth Scientific Industrial Research Organization (CSIRO; a publicly funded government research organization), Science Industry Endowment Fund, National Health and Medical Research Council of Australia (project grant 1011689), Alzheimer's Association, Alzheimer's Drug Discovery Foundation, and an anonymous foundation. See www.aibl.csiro.au for further details.

The ICBM data used in the preparation of this manuscript were supported by Human Brain Project grant PO1MHO52176-11 (ICBM, P.I. Dr John Mazziotta) and Canadian Institutes of Health Research grant MOP- 34996.

The IXI data used in the preparation of this manuscript were supported by the U.K. Engineering and Physical Sciences Research Council (EPSRC) GR/S21533/02 - http://www.brain-development.org/.

The ABIDE data used in the preparation of this manuscript were supported by ABIDE funding resources listed at http://fcon_1000.projects.nitrc.org/indi/abide/. ABIDE primary support for the work by Adriana Di Martino was provided by the NIMH (K23MH087770) and the Leon Levy Foundation. Primary support for the work by Michael P. Milham and the INDI team was provided by gifts from Joseph P. Healy and the Stavros Niarchos Foundation to the Child Mind Institute, as well as by an NIMH award to MPM (R03MH096321). http://fcon_1000.projects.nitrc.org/indi/abide/

This manuscript reflects the views of the authors and may not reflect the opinions or views of the database providers.